\documentclass[
  aps,
  prl,
  reprint,
  superscriptaddress,
  amsmath,
  amssymb,
  floatfix
]{revtex4-2}

\usepackage{graphicx}
\usepackage{bm}
\usepackage{xcolor}
\usepackage{hyperref}
\usepackage{microtype}
\usepackage{placeins}
\usepackage{needspace}
\hypersetup{hidelinks}

\newcommand{\ii}{\mathrm{i}}
\newcommand{\phistar}{\phi_\star}

\newenvironment{finalrevblock}{}{}
\newcommand{\hrrev}[1]{#1}
\newenvironment{hrrevblock}{}{}
\newcommand{\codexrev}[1]{#1}

\newcommand{\codexfinal}[1]{#1}
\newenvironment{codexfinalblock}{}{}
\newcommand{\hrcomment}[1]{#1}

\begin{document}

\title{Josephson-Phase Reversal of Non-Bloch Andreev Propagation}

\author{X. Z. Zhang}
\affiliation{College of Physics and Materials Science, Tianjin Normal University, Tianjin 300387, China}
\affiliation{Interdisciplinary Center, Tianjin Normal University, Tianjin 300387, China}

\author{H. C. Ren}
\email{ren@tju.edu.cn}
\affiliation{\hrrev{\hrcomment{Center for Joint Quantum Studies \& Tianjin Key Laboratory of Low Dimensional Materials Physics and Preparing Technology, Department of Physics, School of Science, Tianjin University, Tianjin 300072, China}}}

\begin{abstract}
Non-Hermitian control is difficult in solid-state systems due to fixed dissipation.
Here, we show that in a spin--orbit-coupled planar Josephson junction, changing only the Josephson phase reverses the non-Bloch propagation of a low-energy Andreev band and switches its boundary accumulation between junction edges.
The phase reshapes the band's spin and electron--hole composition, causing a fixed reservoir to unequally attenuate counterpropagating modes.
Weak-loss theory links this phase-controlled loss imbalance to boundary-selected complex momenta, offering an in situ route to reconfigurable non-Hermitian transport in superconducting platforms.
\end{abstract}

\maketitle

\begin{finalrevblock}
\textit{Introduction.---}\ Non-Hermitian physics has shown that dissipation can do more than broaden quantum levels: it can reshape spectra, topology, and wave propagation
\cite{Bender1998,Heiss2012,ElGanainy2018,MiriAlu2019,Ashida2020,Kawabata2019,Bergholtz2021}.
\hrrev{A striking example is the non-Hermitian skin effect, in which bulk modes under open boundaries become exponentially concentrated toward one boundary}
\cite{HatanoNelson1996,HatanoNelson1997,Lee2016,YaoWang2018,Kunst2018}.
\hrrev{Non-Bloch band theory captures this boundary-sensitive propagation through complex momenta selected by the boundary problem}
\cite{YaoSongWang2018,Yokomizo2019,Okuma2020,Borgnia2020,ZhangYangFang2020,Gong2018}.
\hrrev{Controlling this response in situ, however, remains difficult in solid-state quantum devices, where loss channels are largely fixed by materials, interfaces, and contacts and cannot be tuned independently of the coherent Hamiltonian.}
\hrrev{The central challenge is therefore to reverse a non-Bloch spatial response without rebuilding the dissipative structure that creates it.}

\hrrev{Reservoir engineering has shown that selective dissipation can prepare quantum states and direct their motion}
\cite{MetelmannClerk2015,Verstraete2009,Diehl2008},
\hrrev{while boundary accumulation can arise even without gain}
\cite{Longhi2020,Helbig2020,Weidemann2020}.
\hrrev{In most realizations, however, the direction of this spatial bias is encoded in asymmetric couplings, a designed gain--loss profile, or the system--environment coupling itself.}
\hrrev{Reversing the accumulation therefore requires re-engineering the very structure that produces it.}
\hrrev{This raises a basic question: can non-Bloch propagation be reversed \hrcomment{by a purely Hermitian control parameter, without changing the reservoir or relying on a Hermitian gap closing}?}

Phase-biased Josephson junctions provide a natural setting for realizing this possibility.
\hrrev{Their Andreev states are coherently controlled by the superconducting phase difference and have a phase-tunable spin and electron--hole composition}
\cite{BeenakkerVanHouten1991,Beenakker1991,Bretheau2013,Tosi2019,Hays2021,Nichele2020}.
\hrrev{In extended junctions, spin--orbit coupling and multichannel confinement turn these states into propagating Andreev bands along the weak link}
\cite{ParkYeyati2017,Suominen2017,Pientka2017,Hell2017,Ren2019,Fornieri2019}.
\codexfinal{Exceptional structures have been studied in non-Hermitian systems with particle--hole symmetry and in non-Hermitian superconductors}
\cite{OkugawaYokoyama2019,Tamura2022,Zhu2023}.
\codexfinal{More recently, phase-biased non-Hermitian Josephson junctions have been investigated directly}
\cite{Klees2024,LiTrauzettel2025}.
\codexfinal{Josephson phases also control spin-dependent supercurrents}
\cite{Buzdin2008,Reynoso2008}
\codexfinal{and topological transitions in multiterminal junctions}
\cite{Riwar2016,MeyerHouzet2017,Xie2017}.
\hrrev{Here we show that the same phase can instead reverse a dissipative spatial response.}
\hrrev{In a spin--orbit-coupled planar junction with a fixed spin-selective reservoir, changing only the phase difference reverses the non-Bloch propagation of a low-energy Andreev band and transfers its boundary accumulation from one end of the junction to the other.}
\hrrev{The phase reshapes the spin and electron--hole composition of the band, thereby reversing the relative attenuation of its counterpropagating components by the same reservoir.}
\hrrev{This loss imbalance is converted into boundary-selected complex momenta and opposite finite-chain density profiles, while the electron--hole separation remains finite and the tracked band stays isolated from the next positive-energy band.}
\hrrev{The Josephson phase therefore controls not only Andreev spectra and supercurrents, but also the direction and energy window of dissipative quasiparticle propagation, providing an in situ route to reconfigurable non-Hermitian transport in an established solid-state platform.}

\end{finalrevblock}

\begin{figure}[!t]
\includegraphics[width=\columnwidth]{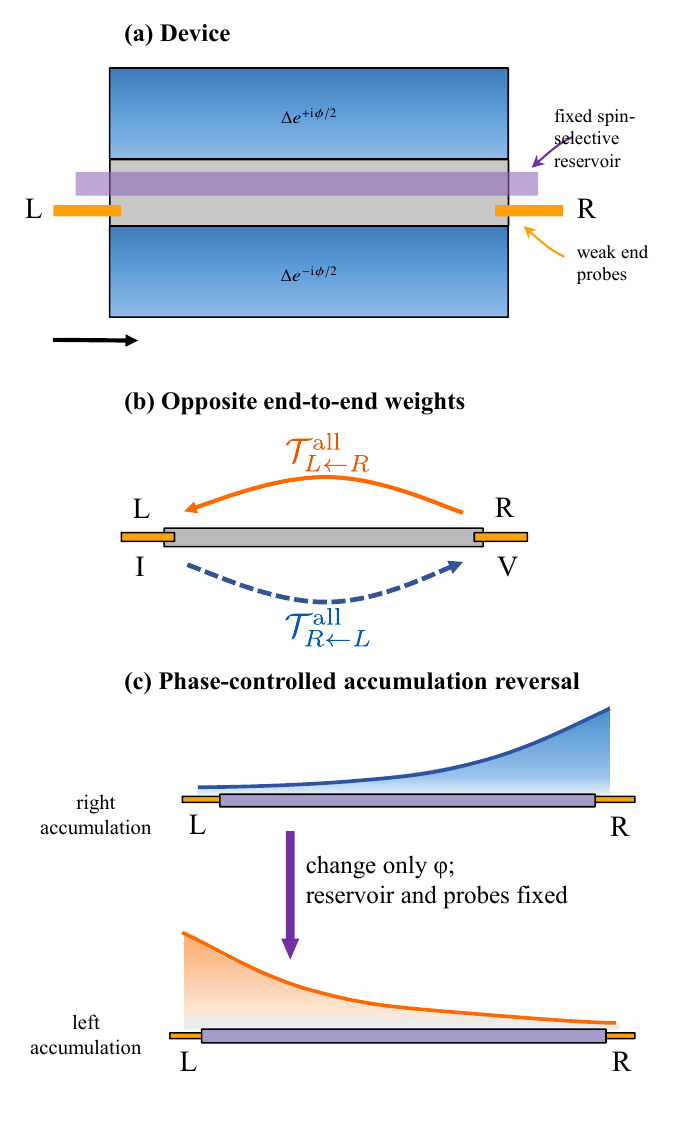}
\caption{
\hrrev{
\label{fig:concept}
Phase-only reversal of Andreev accumulation with a fixed reservoir.
(a) A planar Josephson junction consists of superconducting banks with pair potentials
\(\Delta e^{\pm i\phi/2}\), a spin--orbit-coupled weak link, a fixed spin-selective reservoir, and weak probes at the two ends.
The probes serve only as readout channels and do not enter \(H_{\rm eff}\).
(b) \codexfinal{Opposite complete end-to-end weights \(\mathcal T^{\rm all}_{L\leftarrow R}\) and \(\mathcal T^{\rm all}_{R\leftarrow L}\) read by the same phase-independent weak probes. The probes couple to the complete multilevel Green function, so the other BdG levels remain present; within the fixed spin-filtered window quantified below, the complete outgoing-root ratio reverses which propagation direction is stronger.}
(c) Sweeping only \(\phi\), while the reservoir and probes remain fixed, transfers the boundary accumulation of the same Andreev band from the right end for \(\phi<\phi_\star\) to the left end for \(\phi>\phi_\star\).
With \(\rho_A^R(x)\propto e^{-2\kappa_Ax}\), the two regimes correspond to \(\kappa_A<0\) and \(\kappa_A>0\), respectively.
Quantitative open-boundary profiles are shown in Fig.~\ref{fig:gbz_obc}.
}}
\end{figure}
\begin{hrrevblock}
\textit{Model and directional decay.---}\ The planar Josephson junction in Fig.~\ref{fig:concept} consists of a spin--orbit-coupled weak link extending along \(x\) between two superconducting banks with phases \(+\phi/2\) and \(-\phi/2\).
A fixed spin-selective reservoir couples locally to the weak-link region.
\end{hrrevblock}
\Needspace{6\baselineskip}
The retarded effective Bogoliubov--de Gennes (BdG) Hamiltonian is
\begin{equation}
\begin{aligned}
H_{\rm eff}(k,\phi)
&=
H_{\rm BdG}(k,\phi)
-\ii f_{\rm res}(y)
\left[
\Gamma_0\tau_0
+
\gamma\,\bm m\cdot\bm{\sigma}\tau_z
\right].
\end{aligned}
\label{eq:Heff}
\end{equation}
\begin{hrrevblock}
\codexfinal{Here \(f_{\rm res}(y)\) is the transverse reservoir profile, \(\Gamma_0\) and \(\gamma\) are its scalar and spin-selective decay parameters, and \(\bm m\) is its fixed spin axis; \(\bm{\sigma}\) acts in spin space and \(\tau_z\) in electron--hole space.}
All results below use \(\bm m=\hat{\bm y}\).
The reservoir is uniform along \(x\), introduces no asymmetric longitudinal hopping, and remains fixed throughout the phase sweep; \(\Gamma_0\ge|\gamma|\) ensures passivity.
The microscopic model and the cancellation of direction-independent scalar loss are derived in the \hrcomment{Supplemental Material~\cite{SM}}, Secs.~S1.1--S2.3.
\end{hrrevblock}

\begin{hrrevblock}
We continuously track one positive-energy Andreev band \(A\) by wave-function overlap rather than by energy ordering.
\codexrev{\hrcomment{Its separation from the next positive-energy band, \(\Delta_{\rm iso}(k,\phi)=E_{\rm next}^{+}(k,\phi)-E_A^{+}(k,\phi)\), remains finite over the parameter range considered.}}
The continuation procedure and the distinction between \(\Delta_{\rm iso}\) and the electron--hole separation are documented in the Supplemental Material~\cite{SM}, Secs.~S4.1 and S4.2.
At a fixed real energy, this band has two counterpropagating states, \(|A,R\rangle\) and \(|A,L\rangle\), with signed velocities \(v_R>0\) and \(v_L<0\).
\codexfinal{We use \(\kappa_A\) generically for the selected-band spatial exponent and distinguish its GBZ and fixed-real-energy realizations below; \(\kappa_A^{\rm pert}\) denotes the leading weak-loss form of the latter.}
\codexfinal{To organize the expansion, the two reservoir decay parameters are scaled together as}
\((\Gamma_0,\gamma)\mapsto\lambda(\Gamma_0,\gamma)\).
\end{hrrevblock}

\hrrev{To leading order in \(\lambda\), the spatial exponent is}
\begin{equation}
\begin{aligned}
\frac{\kappa_A^{\rm pert}(k,\phi)}{\gamma}
&=
\frac{1}{2}
\left(
\frac{o_R}{v_R}
+
\frac{o_L}{v_L}
\right)
+O(\lambda),
\\
o_\eta
&=
\langle A,\eta|
O_{\rm loss}
|A,\eta\rangle,
\qquad
\eta=R,L .
\end{aligned}
\label{eq:kappa_pert}
\end{equation}

\begin{hrrevblock}
Here
\(
O_{\rm loss}
=
f_{\rm res}(y)\bm m\cdot\bm{\sigma}\tau_z
\)
measures the spin-selective reservoir weight of each counterpropagating state.
With \(x\) increasing from left to right and
\(\rho_A^R(x)\propto e^{-2\kappa_Ax}\),
positive and negative \(\kappa_A\) correspond to accumulation toward the left and right ends, respectively.
\codexfinal{Equation~\eqref{eq:kappa_pert} converts the signed spin-selective contribution \(\gamma o_\eta\) to temporal decay into spatial attenuation through the signed velocity \(v_\eta\).}
\codexfinal{In the clean mirror-related geometry, the scalar reservoir weights
\(w_\eta=\langle A,\eta|f_{\rm res}|A,\eta\rangle\)
obey \(w_L=w_R\) and cancel because \(v_L=-v_R\), whereas \(o_L=-o_R\), so the spin-selective contributions reinforce one another after division by the oppositely signed velocities.}
\codexfinal{The fixed-real-energy complex-momentum expansion and its validity conditions are given in the Supplemental Material~\cite{SM}, Secs.~S2.1--S2.3.}
No asymmetric hopping, directional boundary condition, or phase-dependent reservoir is introduced.
\end{hrrevblock}

\begin{finalrevblock}

\begin{finalrevblock}

\begin{hrrevblock}
\textit{Why the direction reverses.---}\ Near \(k=0\), the clean mirror-related geometry gives
\(E_A(k,\phi)=E_A(0,\phi)+E_A''(\phi)k^2/2+\cdots\)
and
\(\langle O_{\rm loss}\rangle_{A,k}
=o_0(\phi)+o_1(\phi)k+\cdots\).
The even-in-momentum term \(o_0\) is common to the two counterpropagating states and therefore drops out of the directional exponent.
The odd coefficient \(o_1\) instead measures how the spin-selective reservoir weight changes when the Andreev state begins to propagate.
\end{hrrevblock}

\hrrev{This momentum-induced change is generated by the longitudinal velocity operator}
\begin{equation}
V_x
\equiv
\left.
\frac{\partial H_{\rm BdG}(k,\phi)}
{\partial k}
\right|_{k=0}.
\label{eq:Vx}
\end{equation}

\begin{hrrevblock}
In units with \(\hbar=1\), \(V_x\) is the velocity operator at the band center.
\hrcomment{It generates the first-order change of the Andreev wave function with momentum and therefore acts as the motion vertex that mixes the selected state with other BdG levels.}
In physical terms, a moving Andreev state borrows spin and electron--hole character from the surrounding multilevel spectrum.

We denote the electron--hole conjugate of \(|A\rangle\) by \(|\bar A\rangle\).
Its separation from the selected state,
\(\Delta_{\rm ph}(\phi)
=E_A^{+}(0,\phi)-E_A^{-}(0,\phi)
=2E_A(0,\phi)\),
is distinct from the isolation \(\Delta_{\rm iso}\) from the next positive-energy band.
\end{hrrevblock}

\hrrev{The resulting low-momentum exponent is}
\begin{equation}
\begin{aligned}
\frac{\kappa_A^{\rm pert}}{\gamma}
&=
\frac{o_1}{E_A''},
\\
o_1
&=
2\,\operatorname{Re}
\sum_{l\neq A}
\frac{
\langle A|O_{\rm loss}|l\rangle
\langle l|V_x|A\rangle
}{
E_A-E_l
},
\\
o_1
&\equiv
o_1^{\bar A}
+
o_1^{\rm other}.
\end{aligned}
\label{eq:o1_multiband}
\end{equation}

\begin{hrrevblock}
All states and matrix elements in Eq.~\eqref{eq:o1_multiband} are evaluated at \(k=0\) and at the same Josephson phase.
\hrcomment{Here \(o_1^{\bar A}\) is the contribution from the electron--hole conjugate level \(l=\bar A\), whereas \(o_1^{\rm other}\) is the coherent sum over all remaining positive- and negative-energy BdG levels, \(l\neq A,\bar A\).}

The motion vertex \(V_x\) changes the wave function as the state begins to propagate, while \(O_{\rm loss}\) measures how the borrowed components couple to the fixed reservoir.
The response therefore depends on both energy denominators and phase-dependent matrix elements.
Although the conjugate level has the smallest denominator, the coherent contribution of the remaining levels can acquire the opposite sign as \(\phi\) reshapes the wave functions.

\hrcomment{We define the low-momentum reversal phase by
\(o_1[\phistar(0)]=0\).
Because \(E_A''\) remains finite, this condition also gives
\(\kappa_A^{\rm pert}[0,\phistar(0)]=0\).
At the same phase, both the electron--hole separation \(\Delta_{\rm ph}\) and the isolation \(\Delta_{\rm iso}\) from the next positive-energy band remain finite.
The reversal therefore originates from destructive interference among virtual BdG-level contributions, rather than from a Hermitian gap closing or a switch to another band.}
\codexfinal{Figure~\ref{fig:mechanism} resolves the opposite contributions of the electron--hole conjugate level and all remaining BdG levels. Their cancellation drives the total \(o_1\) through zero while both Hermitian separations remain finite.}
\codexfinal{The derivation, complete-state decomposition, numerical values of \(\Delta_{\rm ph}\) and \(\Delta_{\rm iso}\), and an independent finite-difference check that the curvature remains nonzero are given in the Supplemental Material~\cite{SM}, Secs.~S3.1--S3.3.}
\end{hrrevblock}

\end{finalrevblock}

\begin{figure}[tbp]
\includegraphics[width=\columnwidth]{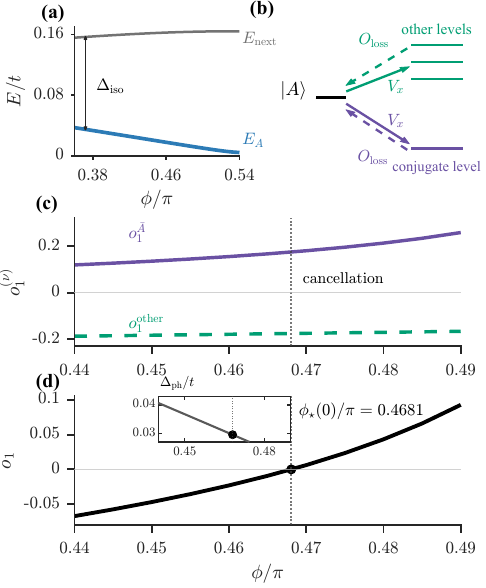}
\caption{
\hrrev{
\label{fig:mechanism}
Multilevel interference reverses Andreev propagation.
(a) The positive-band isolation \(\Delta_{\rm iso}(0,\phi)\) keeps the tracked Andreev band separated from the next positive-energy band at \(k=0\).
(b) The motion vertex \(V_x\) creates virtual admixture with other BdG levels, while \(O_{\rm loss}\) reads how this admixture couples to the fixed reservoir.
(c) The electron--hole conjugate level \(|\bar A\rangle\) and the coherent sum over all remaining BdG levels contribute with opposite signs.
(d) Their cancellation drives \(o_1\) through zero at \(\phistar(0)\); \hrcomment{the inset shows that the electron--hole separation \(\Delta_{\rm ph}/t\) remains finite at the reversal}.
}}
\end{figure}

\begin{finalrevblock}

\begin{hrrevblock}
\textit{From unequal lifetimes to boundary accumulation.---}\ The weak-loss result identifies which counterpropagating state decays more slowly.
Open boundaries convert this decay-rate imbalance into selected complex wave factors and hence a spatial response.

\hrcomment{Under open boundary conditions, the generalized Brillouin zone (GBZ) replaces the real Bloch factor \(e^{\ii ka}\) by a complex factor \(\beta\).
For the selected Andreev band, we denote the two GBZ roots continuously connected to its Hermitian right- and left-moving states by
\(\beta_R^{\rm GBZ}\) and \(\beta_L^{\rm GBZ}\).
Their equal modulus determines the open-boundary bulk solution, whose spatial exponent is}
\end{hrrevblock}
\begin{equation}
\begin{aligned}
|\beta_R^{\rm GBZ}|
&=
|\beta_L^{\rm GBZ}|,
\\
\kappa_A^{\rm GBZ}
&=
-\frac{1}{2a}
\ln
\left|
\beta_R^{\rm GBZ}
\beta_L^{\rm GBZ}
\right|.
\end{aligned}
\label{eq:gbz}
\end{equation}

\begin{hrrevblock}
A common modulus different from unity gives an exponential spatial bias, with the sign of \(\kappa_A^{\rm GBZ}\) selecting the favored end of the junction.
\codexrev{\hrcomment{With \(N_y=20\) transverse sites and four spin--Nambu components per site, the nearest-neighbor problem is a quadratic eigenvalue equation of dimension \(M=80\) and therefore yields \(2M=160\) complex-momentum roots.}}
These roots are first ordered globally by modulus.
The open-boundary bulk condition is imposed on the middle pair, and wave-function continuation from the Hermitian right- and left-moving states verifies that this pair belongs to the selected Andreev band.
The construction is detailed in the Supplemental Material~\cite{SM}, Secs.~S5.1 and S5.2.

\codexfinal{Figure~\ref{fig:gbz_obc} connects this decay imbalance to boundary accumulation: the GBZ roots and the fixed-real-energy complex-momentum roots determine the spatial exponents, while a finite open chain displays the density transfer between the two ends.}
The GBZ and real-energy exponents need not vanish at exactly the same phase because they impose different spectral conditions.
All three nevertheless change sign within the same narrow interval near \(k_0a=0.08\), together with the finite-chain accumulation.

For the real-space profiles in Fig.~\ref{fig:gbz_obc}(b), the same Andreev band is followed at \(\phi/\pi=0.38\), \(0.4718\), and \(0.52\), with identical reservoir parameters.
The outer phases expose opposite biases, while the middle profile is nearly neutral.
The finite-chain state selection, envelope fitting, length convergence, and wave-function comparison are described in the Supplemental Material~\cite{SM}, Secs.~S6.1--S6.4.
\end{hrrevblock}

\end{finalrevblock}

\begin{figure}[tbp]
\includegraphics[width=\columnwidth]{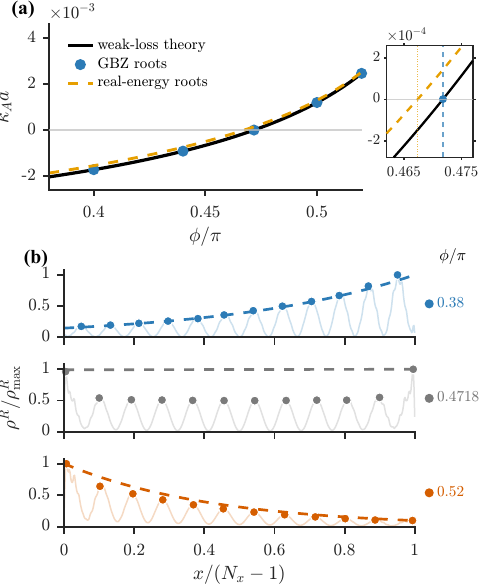}
\caption{
\hrrev{
\label{fig:gbz_obc}
From directional decay to boundary accumulation.
(a) At \(k_0a=0.08\), the weak-loss exponent
\(\kappa_A^{\rm pert}\), the GBZ exponent
\(\kappa_A^{\rm GBZ}\), and the real-energy propagation exponent
\(\kappa_A^{\rm prop}\) all change sign within the same narrow phase interval.
\hrcomment{Inset: magnified view of the three closely spaced reversal points.}
\codexrev{(b) Individually normalized open-boundary right-eigenstate densities \(\rho_A^R(x)\) of the same Andreev band at \(\phi/\pi=0.38\), \(0.4718\), and \(0.52\), with identical reservoir parameters. Symbols mark antinode maxima, and dashed curves show the fitted exponential envelopes. The corresponding half-chain ratios are \(W_R/W_L=2.56\), \(1.00\), and \(0.30\), identifying right accumulation, a nearly neutral profile, and left accumulation, respectively. The commensurate \(N_x=470\) construction and fitting details are given in the Supplemental Material~\cite{SM}, Sec.~S6.4.}
}}
\end{figure}

\FloatBarrier

\begin{codexfinalblock}
\textit{Momentum- and energy-selective reversal.---}\ Away from \(k=0\), the phase at which the weak-loss exponent vanishes disperses with momentum.
We define this reversal line by \(\kappa_A^{\rm pert}[k,\phistar(k)]=0\).
Figure~\ref{fig:front}(a) shows that the same phase does not reverse the entire Andreev band at once: momentum sectors on opposite sides of the zero line favor opposite ends.
At \(\phi/\pi=0.48\), the monotonic band dispersion converts this momentum selectivity into energy windows with opposite propagation bias [Fig.~\ref{fig:front}(b)].
The finite-momentum map uses the full weak-loss expression at each \(k\); the explicit conjugate-level decomposition in Eq.~\eqref{eq:o1_multiband} is invoked only in its controlled low-momentum regime.
\end{codexfinalblock}

\begin{figure}[tbp]
\includegraphics[width=\columnwidth]{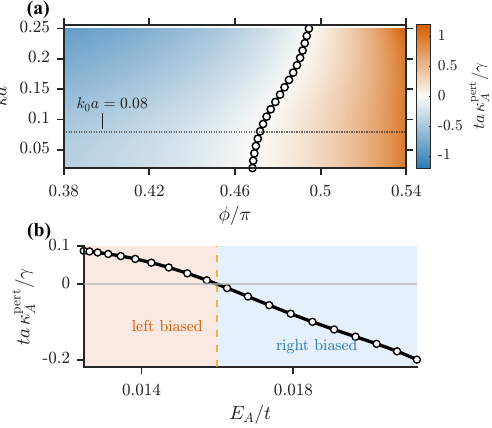}
\caption{\codexfinal{
\label{fig:front}
Momentum- and energy-selective reversal of Andreev propagation.
(a) The weak-loss exponent versus phase and momentum. Black open circles give the zero-line extraction \(\phistar(k)\), and the dotted line marks \(k_0a=0.08\).
(b) At \(\phi/\pi=0.48\), the same Andreev band contains energy sectors with opposite signs of \(\kappa_A^{\rm pert}\) and hence opposite propagation bias.
The full finite-momentum calculation and zero-line extraction are described in the Supplemental Material~\cite{SM}, Sec.~S5.4.}}
\end{figure}

\begin{codexfinalblock}
\textit{Experimental access.---}\ A tunnel-coupled metallic ferromagnetic strip can provide the fixed spin-dependent escape channel, while weak probes at the two ends read the resulting propagation.
Local tunneling spectroscopy locates the selected Andreev resonance, its linewidth, and its separation from neighboring resonances.
Directionality is then identified by the off-diagonal end-to-end response rather than by a local spectrum alone.
\codexfinal{For one fixed spin-polarized electron channel, the selected-band root dominates over a finite phase interval, while the stronger of the two complete positive channel-resolved end-to-end responses switches with phase.}
\codexfinal{The finite-distance root formula, complete multilevel comparison, and weak-probe conditions are given in the End Matter and the Supplemental Material~\cite{SM}, Sec.~S5.5.}
\end{codexfinalblock}

\begin{hrrevblock}
\textit{Conclusion.---}\ A fixed reservoir does not by itself determine a unique direction of propagation; the response also depends on the wave function on which it acts.
In the Josephson junction, the superconducting phase reshapes the spin and electron--hole composition of an Andreev band, changing the interference among different BdG levels until the directional decay imbalance reverses sign.
Boundary-selected complex momenta and finite-chain densities translate this microscopic sign change into a transfer of accumulation between the two ends, while its momentum dependence creates energy windows with opposite propagation bias.
The \hrcomment{overall principle} extends beyond the specific Andreev spectrum considered here: coherent control can reconfigure reservoir-induced propagation by reshaping the multilevel state on which a fixed environment acts.
\hrcomment{The Josephson phase therefore extends its role beyond Andreev spectra and supercurrents to in situ control of both the direction and the energy window of non-Bloch quasiparticle propagation, bridging coherent quantum-phase control and non-Bloch transport in an established solid-state platform.}
\end{hrrevblock}

\begin{acknowledgments}
X. Z. Zhang acknowledges support from the National Natural Science Foundation of China under \codexfinal{Grant Nos.~12675024 and 12275193}.
\hrrev{\hrcomment{This work was also supported by the Fundamental and Interdisciplinary Disciplines Breakthrough Plan of the Ministry of Education of China under Project No.~JYB2025XDXM410.}}
\end{acknowledgments}

\par
\begingroup
\leftskip=0pt
\rightskip=0pt
\parfillskip=0pt plus 1fil
\parindent=1em
\textit{Data availability.---}The data and code supporting this study are available from the corresponding author upon reasonable request.\par
\endgroup
\end{finalrevblock}

\section*{End Matter}

\begin{codexfinalblock}
\textit{From complex momentum to a finite-distance signal.---}\ At a real probe energy \(E\), the outgoing complex-momentum roots continuously connected to the Hermitian right- and left-moving states have Bloch factors \(\beta_R^{\rm prop}\) and \(\beta_L^{\rm prop}\).
Their product defines the branch propagation exponent
\(\kappa_A^{\rm prop}=-(2a)^{-1}\ln|\beta_R^{\rm prop}\beta_L^{\rm prop}|\).
For a simple complex-momentum root, propagation between two endpoints separated by \(L\) is the product of a local residue and a factor \(\beta^{L/a}\).
Squaring the amplitudes and taking the ratio of the two directions therefore separates the finite-distance response into a length-independent end factor and a bulk exponential,
\end{codexfinalblock}
\begin{equation}
\mathcal R_A(L,\phi)
=
C_A(E,\phi)
\exp\!\left[
4\kappa_A^{\rm prop}(E,\phi)L
\right],
\label{eq:propagation_ratio}
\end{equation}
\begin{codexfinalblock}
Here \(C_A\) is the ratio of the selected-root weights sampled by identical phase-independent end probes.
The factor \(4\) has a direct origin: squaring each root contribution doubles its spatial exponent, and comparing the two opposite propagation directions combines their logarithmic attenuations.
The exponent records the directional attenuation accumulated through the junction, whereas \(C_A\) records how the two ends couple locally to the same selected root and does not grow with \(L\).
Thus \(\mathcal R_A>1\) (\(<1\)) favors the left (right) end, and increasing \(L\) makes the complex-momentum contribution progressively dominant.
For \(L=40a\), \(\mathcal R_A\) crosses unity at \(\phi/\pi\simeq0.47055\), close to the root-only value \(0.46739\).
The small offset is the finite, length-independent residue contribution rather than a second reversal mechanism.
\codexfinal{These \(L=40a\) values use the spin-summed normal-region projector of Sec.~S5.3. The complete fixed-channel test below instead uses a localized \(\sigma_x=-1\) electron projector and \(L=10a\); the two readouts therefore need not cross at exactly the same phase.}

\textit{Multilevel mechanism and fixed-probe readout.---}\ The reversal is generated by virtual admixture with the other BdG levels, but this does not require all of their resonances to be simultaneously resolved by the probes.
At fixed real energy, let
\(G_0^R(E)=[E+\ii0^+-H_{\rm eff}]^{-1}\)
denote the intrinsic retarded Green function before adding the readout contacts.
Its endpoint block is a coherent sum over all outgoing complex-momentum roots and can be separated into the root continuously connected to band \(A\) and the sum of all remaining outgoing roots.
This fixed-energy root decomposition is distinct from the energy-pole form used near a spectrally isolated resonance.
The other BdG levels reshape the wave function of band \(A\) through virtual processes, while energy resolution allows the associated resonance to dominate the measured response.
The mechanism is therefore multilevel even when the readout is centered on one Andreev resonance.

To verify the reversal without adapting a probe to the evolving eigenstate, we use the same local electron channel with \(\sigma_x=-1\), fixed endpoint positions, and the same weak linewidth throughout the phase sweep.
With \(\bm\Gamma_{\alpha}^{\rm p}\) denoting this fixed probe coupling at end \(\alpha=L,R\), the complete positive end-to-end weights and their directional ratio are
\end{codexfinalblock}
\begin{codexfinalblock}
\begin{equation}
\begin{aligned}
\mathcal T^{\rm all}_{\alpha\leftarrow\beta}(E_A)
&=
\operatorname{Tr}\!\left[
\bm\Gamma_{\alpha}^{\rm p}G_0^R(E_A)
\bm\Gamma_{\beta}^{\rm p}G_0^A(E_A)
\right]
+O(\nu_{\rm p}^3),
\\
\mathcal R_{\rm all}
&=
\frac{\mathcal T^{\rm all}_{L\leftarrow R}}
{\mathcal T^{\rm all}_{R\leftarrow L}}.
\end{aligned}
\label{eq:fixed_probe_response}
\end{equation}
\end{codexfinalblock}
\begin{codexfinalblock}
Equation~\eqref{eq:fixed_probe_response} is the leading weak-probe result: each \(\bm\Gamma_\alpha^{\rm p}\) is proportional to the probe linewidth \(\nu_{\rm p}\), while probe-induced corrections to \(G_0^{R,A}\) enter one order higher.
Each \(\mathcal T^{\rm all}\) contains \codexfinal{the root associated with the selected Andreev band}, all other outgoing roots, and their coherent interference.
\codexfinal{Across a phase grid of spacing \(0.005\) over \(0.46\leq\phi/\pi\leq0.52\), the same fixed channel at either inner site of the four-site weak link and \(L=10a\) remains selected-root dominated in both directions: \(\epsilon_{\alpha\beta}\leq0.109\), and the pure-background positive weight is at most \(1.2\%\) of the selected-root weight.}
Over this interval \(\mathcal R_{\rm all}\) increases from \(0.965\) to \(1.099\) and crosses unity at \(\phi/\pi=0.48318\), establishing the reversal with a phase-independent probe.
The reversal therefore persists over a finite interval in which the probe channel remains unchanged and the selected resonance remains identifiable; it is not tied to a single fine-tuned phase point.
The three descriptions now form one physical sequence: virtual transitions through the multilevel spectrum reshape the selected Andreev state, its complex-momentum roots determine the spatial bias accumulated over distance, and fixed end probes convert that bias into an exchange of the two directional signals.
\codexfinal{Neither the reservoir nor the probe couplings are changed during the phase sweep. The readout is evaluated at the tracked resonance energy \(E_A(\phi)\), while \(\phi\) is the only parameter that changes the junction Hamiltonian.}
The result above is the positive channel-resolved weight appropriate to this weak-probe test.
A device-specific signed charge conductance further combines electron and hole conversion channels with explicit lead self-energies.
\codexfinal{Details of the Green-function construction and weak-probe conditions are given in the Supplemental Material~\cite{SM}, Sec.~S5.5.}
\end{codexfinalblock}

\FloatBarrier

\end{document}